\documentclass[11pt]{article}

\usepackage{times}
\usepackage{graphicx}

\title{The Mermin Fixed Point}
\author{Veit Elser \\ Department of Physics, Cornell University \\ Ithaca, NY 14853-2501 \\ USA}

\begin{document}

\maketitle

\begin{center}
\parbox{4in}{
The most efficient known method for solving certain computational problems
is to construct an iterated map whose fixed points are by design the problem's
solution. Although the origins of this idea go back at least to Newton, the clearest
expression of its logical basis is an example due to Mermin. A contemporary application
in image recovery demonstrates the power of the method.} 
\end{center}

\section{INTRODUCTION}
Fixed points arise naturally in the study of physical phenomena, be they invariants 
with respect to time evolution (dynamical systems) or rescaling (the renormalization group).
On the practical side, fixed points form the basis of iteration schemes specifically engineered to solve particular
computational problems. One of the simplest applications of this idea is Mermin's solution$^{(1)}$
of the {\it self-referential digit counting puzzle}:

\begin{center}
In this sentence,\\
the digit 0 appears $\underline{\qquad}$ times;\\
the digit 1 appears $\underline{\qquad}$ times;\\
the digit 2 appears $\underline{\qquad}$ times;\\
the digit 3 appears $\underline{\qquad}$ times;\\
the digit 4 appears $\underline{\qquad}$ times;\\
the digit 5 appears $\underline{\qquad}$ times;\\
the digit 6 appears $\underline{\qquad}$ times;\\
the digit 7 appears $\underline{\qquad}$ times;\\
the digit 8 appears $\underline{\qquad}$ times;\\
the digit 9 appears $\underline{\qquad}$ times.
\end{center}

The object is to fill in all the blanks with decimal numerals to make the statement correct. Our
instinct is to treat this as an exercise in logic. Mermin noted that considerably less
mental energy is required by an iterative procedure, where the iterates are the vectors
of ten integers which fill in the blanks: \mbox{$x=(x_0,\ldots,x_9)$}. Starting with an initial vector,
say \mbox{$x(0)=(1,1,1,1,1,1,1,1,1,1)$}, successive iterates are computed by treating the current
vector as a tentative solution and then counting the actual occurrences of the ten digits; thus
$x(1) = (1,11,1,1,1,1,1,1,1,1)$. When the map encounters a fixed point, the tentative solution
is confirmed to be an actual solution. For the choice of initial vector given above, one finds a
{\it Mermin fixed point} already at iterate $x(3)=(1,11,2,1,1,1,1,1,1,1)$. As this example reveals, Mermin fixed points are a powerful and
insufficiently explored strategy for solving a very broad range of
complex problems
\footnote{
The reader is invited to find another Mermin fixed point of this map. There is also a
2-cycle, which solves the even harder problem of finding two such
sentences, each of which has ``this" replaced by ``that."}.

\section{IMAGE RECOVERY}
When a monochromatic plane wave is weakly and elastically scattered by an object, the resulting
diffracted wave has an angular intensity variation derived from the Fourier modes of the object's
scattering density. In this scenario and many others$^{(2)}$, access to the contrast variations within an
object is available only via the modulus of its Fourier transform. Since phase information is
usually not available, a direct inverse Fourier transform cannot be used to recover the object.
On the other hand, relatively mundane properties of the object such as positivity and size, in combination
with the Fourier modulus, may be sufficient
to ``retrieve" the unknown phases and recover the object.

\begin{figure}
\begin{center}
\scalebox{1.6}{\includegraphics{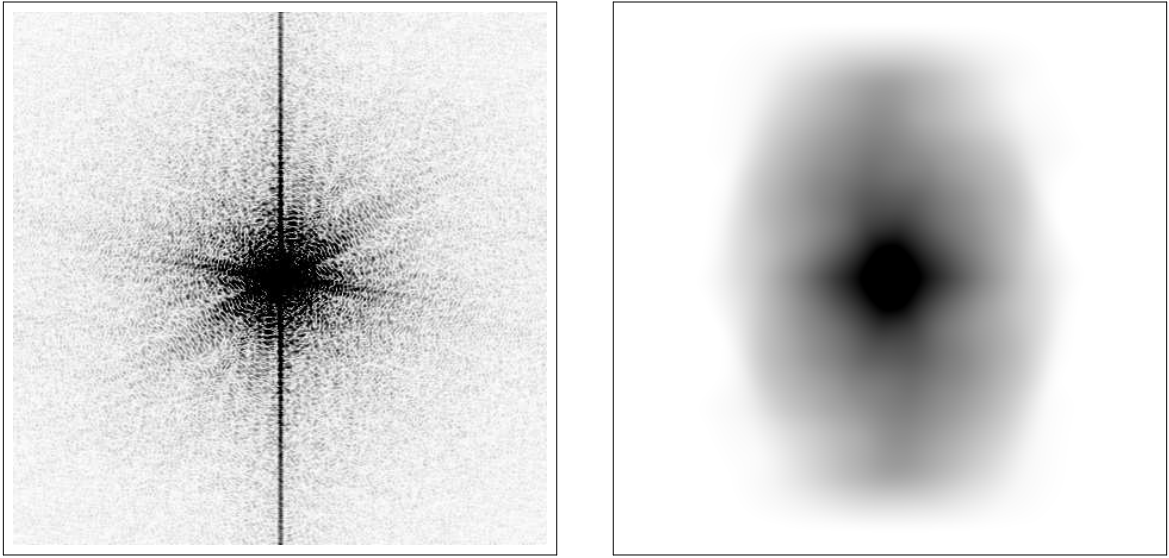}}
\end{center}
\begin{center}
\parbox{5in}{\caption{{\it Left}: Fourier modulus of a two dimensional object such as would be 
produced in a diffraction experiment.
{\it Right}: Autocorrelation of the object obtained by Fourier transforming the squared Fourier modulus.}}
\end{center}
\end{figure}
As an example, consider the speckle intensity pattern shown in Figure 1, formed by 
diffraction from an object in two
dimensions. The data is sampled on an array
measuring $448\times 448$ pixels; we lack the corresponding array of phases,
with which a discrete Fourier transform could be used to directly produce a $448\times 448$ pixel image of the object.

By the convolution theorem, the Fourier transform of the speckle gives the
object's autocorrelation, shown on the right in Figure 1. Close examination of the autocorrelation shows its
support is bounded: specifically, it is negligibly small outside a rectangle measuring
$350\times 440$ pixels. Assuming the object is positive, a well known theorem$^{(3)}$ allows one
to conclude that the
object's support is bounded by a rectangle measuring half the dimensions of
the bound on the autocorrelation support, or $175\times 220$ pixels. This bound on the object's
support, we will see, is sufficient additional information to recover an image of the object.

\begin{figure}
\begin{center}
\scalebox{1.6}{\includegraphics{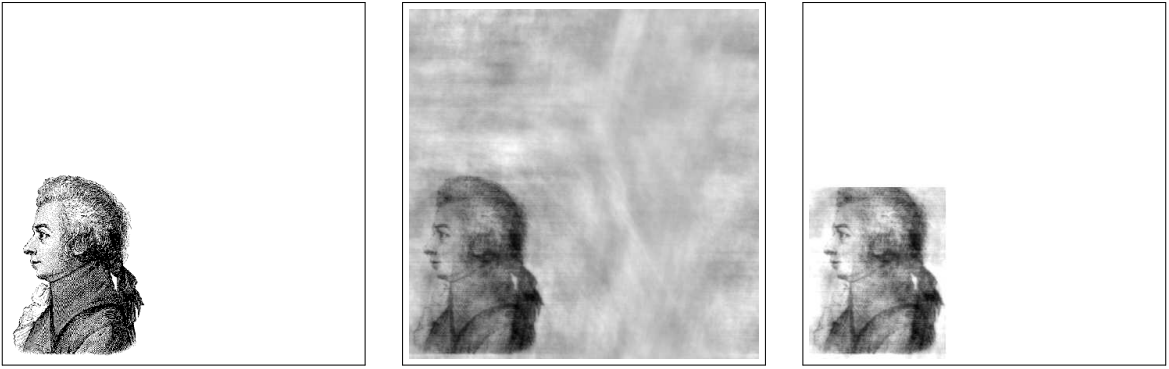}}
\end{center}
\begin{center}
\parbox{5in}{\caption{Projections applied to the arbitrarily chosen image $\rho_{\mathrm{\scriptscriptstyle WA}}$ shown on the left.
{\it Middle}: Fourier modulus projection, 
$\pi_{\mathrm{\scriptscriptstyle F}}(\rho_{\mathrm{\scriptscriptstyle WA}})$. The image shown is
the minimal modification of $\rho_{\mathrm{\scriptscriptstyle WA}}$ such that its Fourier transform
has the modulus shown in \mbox{Figure 1}. {\it Right}: Fourier modulus projection
followed by support/positivity projection, 
\mbox{$\pi_{\mathrm{\scriptscriptstyle S}}\circ\pi_{\mathrm{\scriptscriptstyle F}}
(\rho_{\mathrm{\scriptscriptstyle WA}})$}.}}
\end{center}
\end{figure}

\section{CONSTRAINTS AND PROJECTIONS}
The Fourier modulus data (Fig. 1) and the bound on the object's support are constraints that
the image we are trying to recover must satisfy. 
Given an arbitrary image (for example the image on the left in Figure 2), 
the mathematical operation that restores a
particular constraint, while minimizing the modification of the image, is called a {\it projection}$^{(4)}$. A
natural choice of image metric is the Euclidean metric in the space of pixel values, since it is invariant
with respect to unitary transformations into the Fourier domain. Thus the projection
$\pi_{\mathrm{\scriptscriptstyle F}}$ that restores
the Fourier moduli consists of three operations: (1) transformation of the image to the Fourier domain,
(2) rescaling each complex-valued pixel of the Fourier transform to 
the measured modulus (projection onto a circle), and (3) transformation back to the image domain. Another projection,
$\pi_{\mathrm{\scriptscriptstyle S}}$, restores positivity and the support constraint by setting pixels outside the support,
and negative pixels within the support,
equal to zero. The actions of 
$\pi_{\mathrm{\scriptscriptstyle F}}$ and $\pi_{\mathrm{\scriptscriptstyle S}}$ are illustrated in Figure 2.

Uniqueness in image recovery requires that the number of constraints outnumber the free
variables$^{(5,6)}$. Since a real-valued image of $N$ pixels has approximately $N/2$
unknown phases in its Fourier transform, a bound on the support of the object measuring less than half the
image area is normally sufficient to ensure uniqueness. This condition is easily
satisfied by the example introduced earlier (Fig. 1): there are 100350 independent continuous phases in
a $448\times 448$ pixel Fourier transform, while the bound on the object support
constrains a larger number of pixels, \mbox{$448\times 448 - 175\times 220 = 162204$}, to be zero.

\section{THE DIFFERENCE MAP}
Short of having a projection that directly recovers the image by
simultaneously restoring, from an arbitrary input, both the Fourier modulus {\it and} support/positivity constraints, one can hope to use the projections $\pi_{\mathrm{\scriptscriptstyle F}}$ and $\pi_{\mathrm{\scriptscriptstyle S}}$ in an iterative fashion such
that the solution can be extracted from an appropriate Mermin fixed point. One such approach is the
{\it difference map}$^{(6)}$:
\begin{equation}
\rho\mapsto D(\rho) = \rho + \beta\left[ \pi_{\mathrm{\scriptscriptstyle S}}\circ f_{\mathrm{\scriptscriptstyle F}}(\rho)-\pi_{\mathrm{\scriptscriptstyle F}}\circ f_{\mathrm{\scriptscriptstyle S}}(\rho)
\right]\quad.
\end{equation}
The action of $D$ is to add to the current
iterate $\rho$ the difference of projections (composed with two additional maps) scaled by a parameter $\beta$. 
To see how a Mermin fixed point of $D$, $\rho^\ast$, provides the solution, $\rho_{\rm sol}$, we
observe that $D(\rho^\ast)=\rho^\ast$ implies the difference of projections vanishes. In other words,
the same image, now identified as $\rho_{\rm sol}$, was produced by each of the two projections and
therefore satisfies both sets of
constraints:
\begin{equation}
\pi_{\mathrm{\scriptscriptstyle S}}\circ f_{\mathrm{\scriptscriptstyle F}}(\rho^\ast) = \rho_{\rm sol} = \pi_{\mathrm{\scriptscriptstyle F}}\circ f_{\mathrm{\scriptscriptstyle S}}(\rho^\ast)\quad.
\end{equation}
The maps $f_{\mathrm{\scriptscriptstyle F}}$ and $f_{\mathrm{\scriptscriptstyle S}}$ have so far
not been specified but must be chosen with care in order to make the Mermin fixed point
{\it attractive}. \mbox{Reference 6} makes the choice
\begin{equation}
\begin{array}{rl}
f_{\mathrm{\scriptscriptstyle F}} &= (1+\gamma_{\mathrm{\scriptscriptstyle F}})\pi_{\mathrm{\scriptscriptstyle F}}-\gamma_{\mathrm{\scriptscriptstyle F}}\\
f_{\mathrm{\scriptscriptstyle S}} &= (1+\gamma_{\mathrm{\scriptscriptstyle S}})\pi_{\mathrm{\scriptscriptstyle S}}-\gamma_{\mathrm{\scriptscriptstyle S}}\quad,
\end{array}
\end{equation}
and finds $\gamma_{\mathrm{\scriptscriptstyle F}}=\beta^{-1}, \gamma_{\mathrm{\scriptscriptstyle S}}=-\beta^{-1}$ as optimal
parameter values
\footnote{When $f_{\mathrm{\scriptscriptstyle F}}$ and $f_{\mathrm{\scriptscriptstyle S}}$ are identity maps
($\gamma_{\mathrm{\scriptscriptstyle F}}=\gamma_{\mathrm{\scriptscriptstyle S}}=-1$) the Mermin fixed
point is found to be repulsive$^{(6)}$.}.

The difference map is superior to the naive alternating projection map 
$A=\pi_{\mathrm{\scriptscriptstyle S}}\circ \pi_{\mathrm{\scriptscriptstyle F}}$ because of
stagnation caused by fixed points
of $A$ which do not satisfy the Fourier modulus constraint (and have no simple relationship
to the solution which does, in contrast to eq. 2).
To overcome stagnation, Fienup$^{(7)}$ introduced the {\it hybrid input-output} map which, interestingly, is
obtained as a special case of the difference map
for the parameter
values \mbox{$\gamma_{\mathrm{\scriptscriptstyle F}}=\beta^{-1}$}, 
\mbox{$\gamma_{\mathrm{\scriptscriptstyle S}}=-1$}.
 Although the hybrid input-output
map has been the main tool for phase retrieval for nearly
twenty years, its fixed point properties in the geometrical setting of projections has come to
light only recently$^{(6,8)}$.

\section{A PHASE RETRIEVAL EXAMPLE}

When implemented with the fast Fourier transform, the difference map can be computed in a time
that grows only quasi-linearly with the number of pixels in the image. Although there is as yet no 
comprehensive theory of the
number of iterations required to reach a Mermin fixed point, numerical experiments indicate that
progress toward the solution, measured by the norm of the difference,
\begin{equation}
\epsilon = \| \pi_{\mathrm{\scriptscriptstyle S}}\circ f_{\mathrm{\scriptscriptstyle F}}(\rho)-\pi_{\mathrm{\scriptscriptstyle F}}\circ f_{\mathrm{\scriptscriptstyle S}}(\rho) \|\quad,
\end{equation}
is systematic though not strictly monotone.

\begin{figure}
\begin{center}
\scalebox{1.2}{\includegraphics{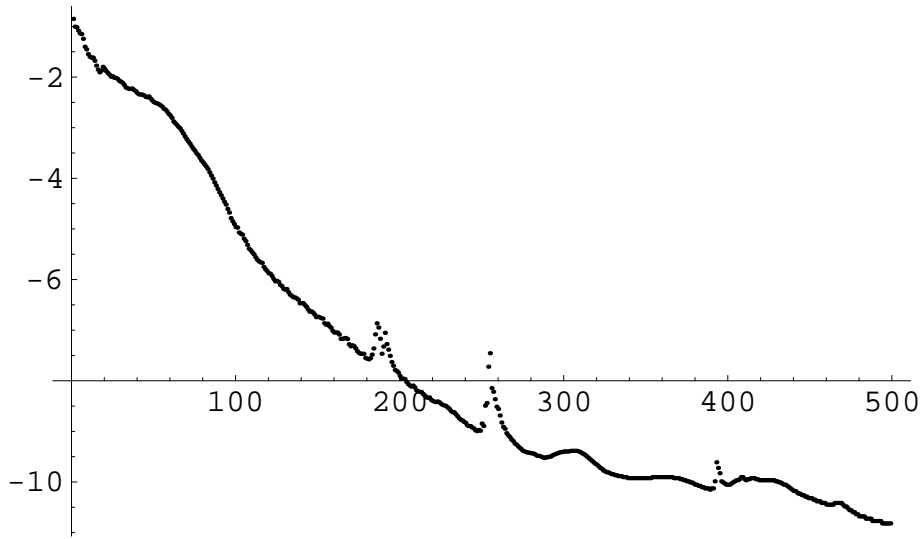}}
\end{center}
\begin{center}
\parbox{5in}{
\caption{Plot of $\log{\epsilon}$ (eq. 4) for the first 500 iterations of the difference map using the diffraction data of Figure 1, the support/positivity constraint as illustrated in Figure 2, and $\beta=1.15$.}}
\end{center}
\end{figure} 

Figure 3 shows the logarithm of the error estimate $\epsilon$ as a
function of iteration for the example introduced in Section 2. A rectangular object support constraint
was imposed, with dimensions ($175\times 220$)
determined from the autocorrelation; $\rho_{\mathrm{\scriptscriptstyle WA}}$ (Fig. 2) was 
the initial image and $\beta=1.15$ was used for all 500 iterations. The evolution of the iterates $\rho(i)$
is shown in Figure 4. The stationarity of the last iterate, together with the smallness of the
corresponding error (Fig. 3), leaves no doubt that a Mermin fixed point has indeed been found.
According to equation 2, the object is recovered by applying the map
$\pi_{\mathrm{\scriptscriptstyle S}}\circ f_{\mathrm{\scriptscriptstyle F}}$ 
(or $\pi_{\mathrm{\scriptscriptstyle F}}\circ f_{\mathrm{\scriptscriptstyle S}}$)
to the final iterate; the result
is shown in Figure 5.
Finally, although the fixed point $\rho(\infty)$ (last image in Fig. 4) is not unique and dependent on the
initial image $\rho(0)$, the uniqueness of the recovered object (Fig. 5)
has convincingly been demonstrated over a period spanning 67 years.
\begin{figure}
\begin{center}
\scalebox{1.6}{\includegraphics{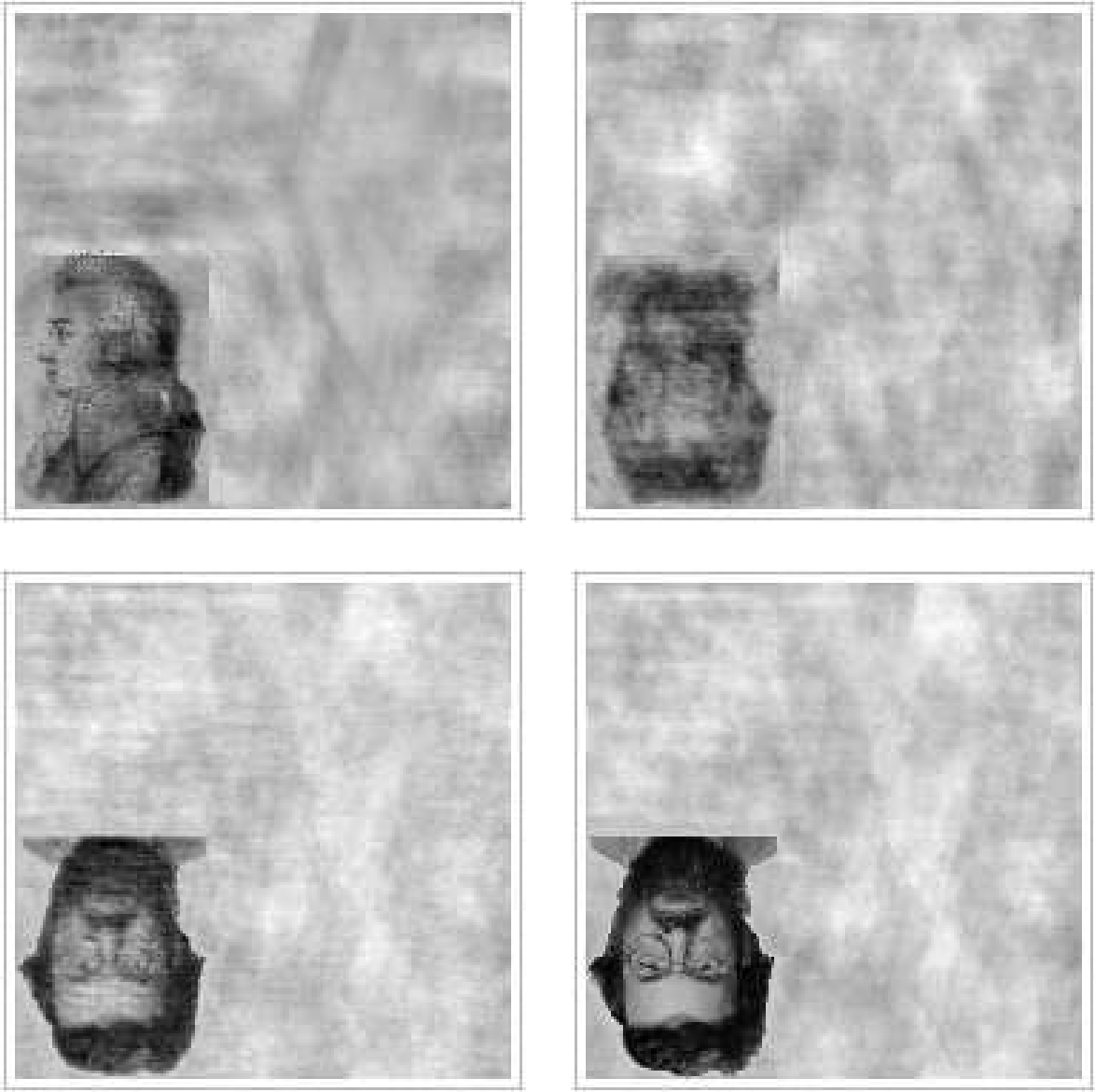}}
\end{center}
\begin{center}
\parbox{5in}{
\caption{Iterates $\rho(1), \rho(5), \rho(50)$ and $\rho(500)$ of the difference map; the initial image was
$\rho(0)=\rho_{\mathrm{\scriptscriptstyle WA}}$ (Fig. 2).}}
\end{center}
\end{figure}

\begin{figure}
\begin{center}
\scalebox{1.6}{\includegraphics{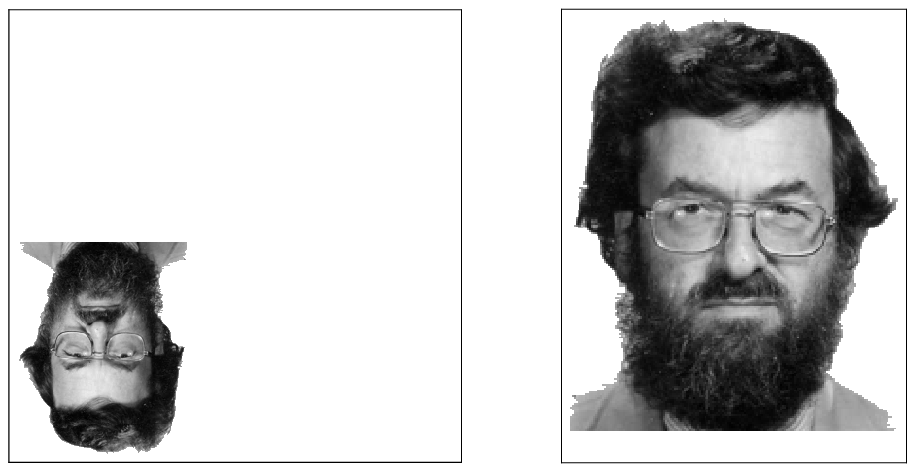}}
\end{center}
\begin{center}
\parbox{5in}{
\caption{The recovered image ($\pi_{\mathrm{\scriptscriptstyle S}}\circ f_{\mathrm{\scriptscriptstyle F}}$ applied to the last image in Figure 4) and detail, after inversion.}}
\end{center}
\end{figure}

\section{CONCLUSIONS}
Mermin's example$^{(1)}$ is the inspiration for iterative solutions
to problems considerably beyond Newton's square root and descendants.
Phase retrieval belongs to the class of {\it feasibility problems}, usually posed
in the context of linear programming with convex constraints$^{(8)}$. Since the Fourier
modulus constraint is nonconvex, standard algorithms either do not apply or have
no guarantee of convergence. The iterative difference map algorithm$^{(6)}$, also without
a bound on the number of iterations, is currently the most efficient method for
solving a large class of phase retrieval problems. This includes
a highly simplified version of phase retrieval: the problem of recovering a fixed length
binary sequence from its cyclic autocorrelation$^{(9)}$. The latter 
appears to be comparable in difficulty, and has a mathematical kinship to,
the problem of factoring integers.
\footnote{One is interested in factoring elements in the ring of cyclotomic integers, with each of the two factors
known {\it a priori} to have binary coefficients and related as algebraic conjugates.}

\section{ACKNOWLEDGMENT}
The author thanks Jason Ho for providing the data (Fig. 1) used in the numerical experiment.
This work was supported by the National Science Foundation under grant ITR-0081775.

\end{document}